\documentclass[prl, twocolumn, amsmath,floatfix, unsortedaddress, superscriptaddress]{revtex4-1}
\usepackage{graphicx}
\usepackage{epstopdf}
\usepackage{natbib}
\usepackage{bm}
\usepackage[T1]{fontenc}
\usepackage{txfonts}
\usepackage{xcolor}

\newcommand{\mfp}{{m\!f\:\!\!p}}
\newcommand{\inlinemaketitle}{{\let\newpage\relax\maketitle}}

\newcommand{\PPPL}{Princeton Plasma Physics Laboratory, Princeton, NJ 08543, USA}
\newcommand{\Princeton}{Department of Astrophysical Sciences, Princeton University, Princeton, New Jersey 08540, USA}
\newcommand{\UofM}{Center for Ultrafast Optical Science, University of Michigan, Ann Arbor, Michigan 48109, USA}
\newcommand{\UofNH}{Space Science Center, University of New Hampshire, Durham, New Hampshire 03824, USA}

\newcommand{\Laser}{Laboratory for Laser Energetics, University of Rochester, Rochester, New York 14623, USA}
\newcommand{\UofW}{Department of Physics, University of Wisconsin-Madison, Madison, Wisconsin 53706, USA}

\begin{document}

\begin{abstract}

First-principles kinetic simulations are used to investigate magnetic field generation processes in expanding ablated plasmas relevant to laser-driven foils and hohlraums.
In addition to Biermann-battery-generated magnetic fields, strong filamentary
magnetic filaments are found to grow in the corona of single expanding plasma plumes; such filaments are observed to dominate Biermann fields at sufficiently large focal radius, reaching saturation values of $\sim$ 100 T at National Ignition Facility-like drive conditions.
The filamentary fields result from the ion Weibel instability driven by relative counterstreaming between the ablated ions and a sparse background population, which could be the result of a gas prefill in a hohlraum or laser pre-pulse.
The ion-Weibel instability is robust with the inclusion of collisions and grows on a timescale of 100 ps, with a wavelength on the scale of 100-250 $\mu$m, over a wide range of background population densities; the instability also gives rise to coherent density oscillations. 
These results are of particular interest to inertial confinement fusion experiments, where such field and density perturbations can modify heat-transport as well as laser propagation and absorption.


\end{abstract}



\title{Weibel instability drives large magnetic field generation in laser-driven single plume ablation} 

\author{J. Matteucci}
\email{jmatteuc@pppl.gov}
\affiliation{\Princeton}
\author{W. Fox}
\affiliation{\Princeton}
\affiliation{\PPPL}
\author{A. Bhattacharjee}
\affiliation{\Princeton}
\affiliation{\PPPL}
\author{D. B. Schaeffer}
\affiliation{\Princeton}
\author{K. Lezhnin}
\affiliation{\Princeton}
\author{K. Germaschewski}
\affiliation{\UofNH}
\author{G. Fiksel}
\affiliation{\UofM}
\author{J. Peery}
\affiliation{\UofW}
\author{S. X. Hu}
\affiliation{\Laser}
\date{\today}

\maketitle

Magnetic fields pervade our entire known universe and fundamentally modify the transport properties of charged particles therein. In laboratory and astrophysical plasmas, there are few mechanisms known to spontaneously generate magnetic fields; one significant mechanism is the Weibel instability, which feeds off of non-equilibrium temperature anisotropy within a plasma, generating magnetic filaments. This instability has been proposed to play a role in the development of astrophysical collisionless shocks where the infrequency of particle collisions requires other mechanisms to mediate shock formation \cite{Sagdeev1966}. More recently, this instability has been observed in the laboratory in several high energy density (HED) laser-driven experiments, where two ablated plasma plumes are driven to collide head on \cite{FoxPRL2013, HuntingtonNatPhys2015}; complementary kinetic computational studies have confirmed the instability growth in such counterstreaming experiments \cite{FoxPRL2013}. The Weibel instability requires large systems ($L \gg d_i$) and significant scale separation between the kinetic ion-skin-depth scale and ion mean-free-path ($\lambda_{\mfp.ii} \gg d_i$), resulting in collisionless ions; such regimes are now readily produced on inertial confinement fusion (ICF) class laser systems such as OMEGA, the National Ignition Facility (NIF), and Laser Megajoule. Therefore the instability has been proposed as a possible explanation \cite{ManuelPoP2013} of anomalous magnetic filamentation observed throughout many non-colliding long-pulse ($\sim$ 1 ns) laser-driven ablation experiments \cite{RyggScience2008, RosenbergPRL2015, GaoPRL2015}; notably, in the context of ICF, such filamentation could disrupt efforts for a symmetric compression necessary for ignition.

\begin{figure*}
\includegraphics[width= 18cm]{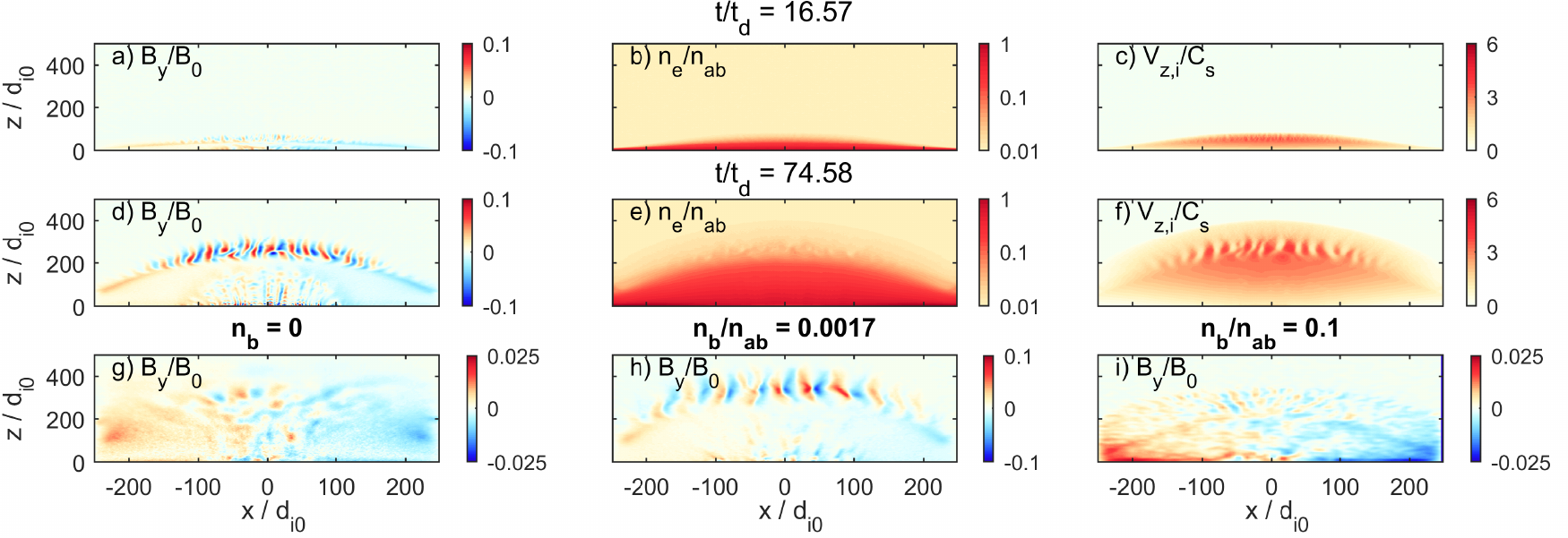}
\caption{(a-c) Out-of-plane magnetic field, electron density, and $V_{zi}$, respectively, at $t/t_{d} =14$, for $n_b/n_{ab}$ = 0.01, $M_i/Z m_e  = 64$. (d-f) Same quantities at $t/t_d = 74.6$. (g-i) Magnetic field at $t/t_d = 75$ for no background density, $n_b/n_{ab}$ = 0.0017, and $n_b/n_{ab}$ = 0.1, respectively; all simulations have 500 particles per cell at $n_{ab}$, where (g-i) are performed at $M_i/Z m_e  = 25$.}
\label{fig:1}
\end{figure*}

In this Letter, we demonstrate via first principles kinetic simulations that the ion Weibel instability grows and generates strong filamentary magnetic fields within the expansion of a single laser-driven plume into a sparse, stationary background plasma. 
We use a novel ablation scheme with the inclusion of particle collisions \cite{FoxPoP2018} to model laser-driven plasma ablation into a low-density background plasma; our setup is matched to the collisionality and scale of laser-driven experiments at OMEGA and NIF where the laser focal radius (several mm) is much larger than the plasma kinetic scales. 
We observe the growth and saturation of filamentary magnetic fields on the order of 25-100 T, outcompeting the Biermann battery effect by a factor of 5$\times$ \cite{HainesPRL1997}. 
The filamentary fields develop within the corona of a single expanding plume where the interaction between ablating ions and stationary background ions is characterized by a long mean free path $\lambda_{mfp,ii}\gtrsim  L$; this allows for relative counterstreaming between the ion populations, driving filamentation. The growth rates and filament wavelengths quantitatively agree with linear ion Weibel theory driven by the measured ion counterstreaming, where growth is strongest in the presence of a background plasma in the range of $0.0017-0.05$ $n_{ab}$, the laser ablation density. 
The kinetic nature of the instability means this growth is fundamentally missed by fluid descriptions of the plasma;
additionally, collisions must be included in the modeling as only counterstreaming ions are effectively collisionless on long pulse experimental timescales; previous collisionless simulations have demonstrated significant Weibel growth due to electron anisotropy in model ablation scenarios \cite{SchoefflerPRL2014, SchoefflerPoP2016}; however these results are more relevant to short-pulse ($\sim$ 1 ps) laser experiments in which electrons are collisionless.
Our results may explain past laser ablation experiments \cite{RyggScience2008, ManuelPoP2013} and have significant implications for on-going experiments; in particular, hohlraums are often filled with helium gas \cite{BerzakPRL2015}, which when ionized would provide a background population in the range for this instability to grow. 
We hypothesize that ion-counterstreaming-driven magnetic filamentation may occur quite generally during plasma ablation; for example, temporal laser intensity modulation, such as a finite ramp time, will result in faster ions streaming through slower ion populations.

\textit{Simulation setup:} To investigate magnetic field generation self-consistently during ablation, we use the PSC code \cite{GermaschewskiJCP2016}, a fully kinetic, explicit particle-in-cell code which includes a binary Monte-Carlo Coulomb collision operator. 2-D simulations are initialized with a flat (in the x-direction), dense target of cold electrons and ions, surrounded by a sparse, cold background plasma. Laser ablation is modeled by heating electrons with a radial profile $H(x) \propto e^{-(x/R_h)^k}$ and periodically replenishing the dense target with cold plasma \cite{FoxPoP2018}. The target density and heating magnitude are tuned to obtain a chosen ablation density $n_{ab}$ and temperature $T_{ab}$ which define the ablation ion skin depth $d_{i0} = (M_i / n_{ab} Z e^2 \mu_0)^{1/2}$ and the sound speed $C_{s, ab} = (Zk_bT_{ab}/M_i)^{1/2}$. This leads to an ablation timescale $t_d = d_{i0}/ C_{s, ab}$ and the characteristic magnetic field $B_0 = \sqrt{\mu_0n_{ab}k_b T_{ab}}$. When normalized to ablation units, the target-normal density evolution agrees well with radiation-hydrodynamic simulations performed with the DRACO code, allowing for a direct conversion from $n_{ab}$ and $T_{ab}$ in PSC to $n_{ab, phys}$ and $T_{ab, phys}$ as measured in radiation-hydrodynamic simulations\cite{SuxingPoP2013}. We model our simulations after recent NIF simulations where $R_h/d_{i0} = 195$  and $k = 4$ at a peak intensity of $1\times 10^{14}$ W/cm$^2$.  From DRACO simulations, $n_{ab, phys}= 4.5\times 10^{27}$m$^{-3}$, $T_{ab, phys} = 1$ keV, $C_{ab, phys} \approx 200$ km/s, and $B_{0, phys} = 1000$ T.  Ref.\ \cite{FoxPoP2018} presents the full process for modeling heating, ablation, and matching to DRACO simulations.

For computational tractability, we use a compressed ratio between $T_{ab}$ and the electron rest mass energy, $m_ec^2/T_{ab}= 25$, and a range of ion-electron mass ratios, $M_i /Zm_e =25-3600$ ($Z=6$ for carbon). We have verified convergence with respect to these parameters and discuss the mass ratio cases below. We use a grid spacing $\Delta x = 0.5 \, d_{e0}$, $2.5 \lambda_{D, ab}$, where $ \lambda_{D, ab} = \sqrt{\epsilon_0k_bT_{ab}/n_{ab}e^2}$. The target density is $2.5 \, n_{ab}$, while the background density ($n_b$) dependence is investigated. All plasma starts at $T_{cold}$ = 0.025 $T_{ab}$. The collisionality, described by $\lambda_{\mfp0}$, the mean free path of electrons at $T_{ab}$ and $n_{ab}$, is matched to the electron skin depth $d_{e0}$ to preserve the correct collisional diffusivity of the magnetic field; $\lambda_{\mfp0}/d_{e0} = 20$. Note, $\lambda_{\mfp0}$ is $\sqrt{(M_i/m_e)_{phys}/(M_i/m_e)_{code}}$ times less collisional on the $d_{i0}$ scale; we investigate this effect with mass ratio studies.

Figure 1(a-c) shows $B_y$, $n_e$, and $V_{zi}$, the ion flow velocity, at an early timestep for a simulation at $M_i/Zm_e = 64$, $n_b/n_{ab} = 0.01$, in a periodic domain of $500 \, \times \, 2000 \, d_{i0}\, (L_x, L_z)$ with 2$\times10^9$ particles. Seen in Fig.\ 1(b,c), as the plume expands up from the target at $z = 0$, it interacts with the local background plasma. In Fig.\ 1(a), on either side of the expansion, the Biermann battery effect generates fields via $\nabla n \times \nabla T$ on the order of $0.02$ $B_0$. Within the center of the expansion, we find the development of magnetic filaments; as found in Ref.\ Schoeffler et al., these filaments develop provided $R_h > \lambda_{w}$, where $\lambda_{w}$ is the filament wavelength.

By $t/t_d = 74.6$, shown in Fig.\ 1(d), the magnetic filamentation has saturated around $0.1\, B_0$, or 100 T in physical units, surpassing the Biermann generation by a factor of 5. The saturation wavelength $\lambda_{w,sat} \approx$ 25 $d_{i0}$, which at the local density corresponds to $\approx 5$ $d_{i, local}$, or 125 $\mu m$. In Fig.\ 1(f) $V_{z,i}$ shows associated filamentation; as seen in Fig.\ 1(e), the instability occurs in the interaction region between the ablation and background plasmas. Density perturbations also grow with the instability, analyzed in detail below. 

Fig.\ \ref{fig:1}(g-h) show the cases of no background plasma, $n_b/n_{ab} = 0.0017$, and $n_b/n_{ab} = 0.1$ at the same time as Fig.\ 1(d), respectively, each at $M_i/Zm_e = 25$.  In Fig.\ 1(g), while there are irregular pockets of magnetic field in the central plume, this field is 10 $\times$ less than observed in Fig.\ 1(d); here the majority of the flux is generated due to Biermann battery effect operating on the sides of the plume. In Fig.\ \ref{fig:1}(h) at $n_b/n_{ab} = 0.0017$ we observe magnetic filaments on the same order as $n_b/n_{ab} = 0.01$, where now the wavelength is larger, on the order of 40 $d_{i0}$, or 200 $\mu m$. In (i), at $n_b/n_{ab} = 0.1$, we observe smaller, 10 $\times$ weaker filamentation. These comparisons indicate (1) there is an optimal range of $n_b$ for this instability to grow, and (2) the instability wavelength depends on $d_{i,local}$.

\begin{figure}
\includegraphics[width= 8.5cm]{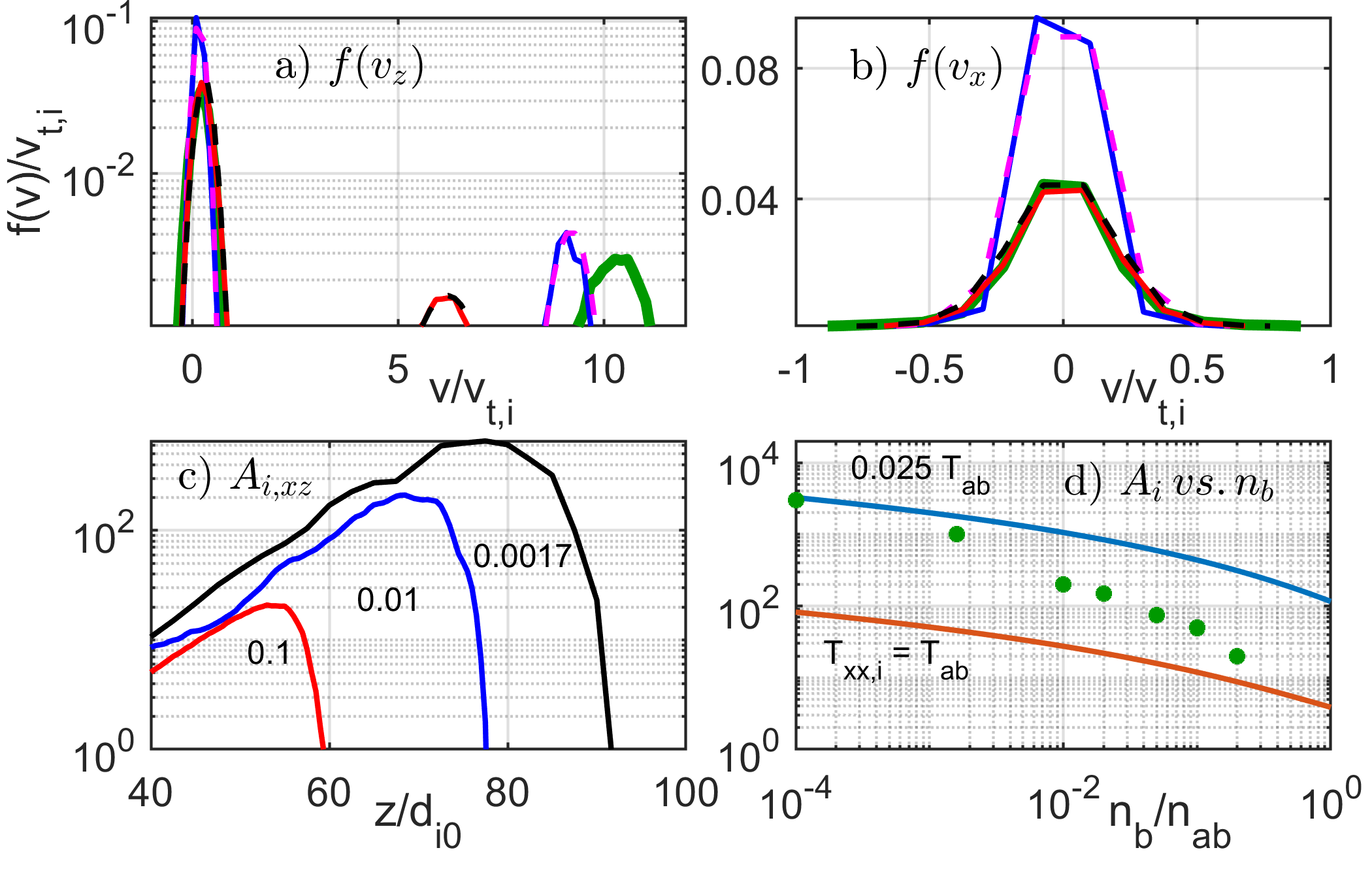}
\caption{(a,b) Ion velocity distribution functions $f(v_z)$ and $f(v_x)$, respectively (where $n_i$ = $\int f(v_j)dv_j$), inside the maximum filamentation at $t/t_d = 14$ for $n_b/n_{ab} = 0.1$ (red), 0.01 (blue), both at $M_i/Z m_e= 25$, and $n_b/n_{ab} = 0.01$, $M_i/Z m_e= 400$ (green). The first two cases are fit in black and magneta. (c) $A_i$ along the $z$ axis for $n_b/n_{ab} = 0.1$ (red), 0.01 (blue), and $n_b/n_{ab} = 0.0017$ (black), all at $M_i/Z m_e= 25$. (d) Model of max($A_i$) vs $n_b$ using $T_{xx,i} = 0.025\, T_{ab}$ (blue) and $T_{zz,i} = T_{ab}$ (red), and green data points from simulations at $M_i/Z m_e= 25$.}
\label{fig:2}
\end{figure}

To study the origin of this instability, we investigate the ion velocity distribution function inside the filamentation. In Fig.\ 2(a,b), $f(v_z)$ and $f(v_x)$ are shown respectively for $n_b/n_{ab} = 0.01$ (blue) and $0.1$ (red) - both at $M_i/Zm_e = 25$. Gaussian fits are shown in magenta and black. In both cases, $f(v_z)$ shows that there are two distinct, counterstreaming ion populations; fits indicate the ablation ions are streaming at roughly 9 $v_{t,i}$ and 6 $v_{t,i}$ for  $n_b/n_{ab} = 0.01, 0.1$ respectively, while the background ions are stationary. Here $v_{t,i} = \sqrt{k_bT_{ab}/M_i} =\frac{ C_{s,ab}}{\sqrt{Z}}$. Conversely, in Fig.\ 2(b), $f(v_x)$ shows that the populations overlap in $v_x$. The relative counterstreaming in $f(v_z)$ yields a large $T_{zz,i}= \int f(v_{z,i})(v_{z,i} - v_{z,0})^2 dv_z$, as compared to $T_{xx,i}$, resulting in a significant ion temperature anisotropy, $A_i  = T_{zz,i}/T_{xx,i} -1$, which is well documented to drive the Weibel instability \cite{DavidsonPoF1972, FoxPRL2013}. 

Physically, the counterstreaming results from the two distinct ion populations. $A_i(x=0,z)$ is plotted in Fig.\ 2(c) for $n_b/n_{ab}$ = 0.0167, 0.01, and 0.1 at $t/t_d = 14$. $A_i$ peaks sooner and at a lower maximum as $n_b$ is increased; this follows from ablation physics \cite{DrakeHED2018}, where the ablated ion density $n_{i, ab}(z) = \frac{n_{ab}}{Z}e^{-z/tC_s}$, with velocity profile $V_z(z) = C_s (1 + z/tC_s)$ and temperature $T_{0}$. Peak anisotropy occurs where $n_{i, ab}(z) \approx n_{b,i}$; for denser $n_b$, this occurs closer to the target at a slower ablation velocity. Superimposing a stationary $n_b$ with temperature $T_0$ on this model, we can solve for max($T_{zz,i}(z)$) at a given $n_b$ and time. Assuming $T_{xx,i}/T_{ab} = 0.025$, the initial temperature, we obtain max($A_i$) vs. $n_b$ at $t/t_{d} = 14$, shown in Fig.\ 2(d) in blue. Similarly, the red line represents max($A_i$) solved with $T_{xx,i}= T_{ab}$, the collisional heating limit for ions heated by ablation electrons; note how in Fig.\ 2(b), at $n_b/n_{ab} = 0.1$ (red), the $f(v_x)$ distribution is noticably hotter, indicating collisional heating. Green dots show max($A_i$) from simulations, indicating $A_i$ is well described by this understanding, provided we account for heating via electrons in denser background cases.

\begin{figure}
\includegraphics[width= 8.5cm]{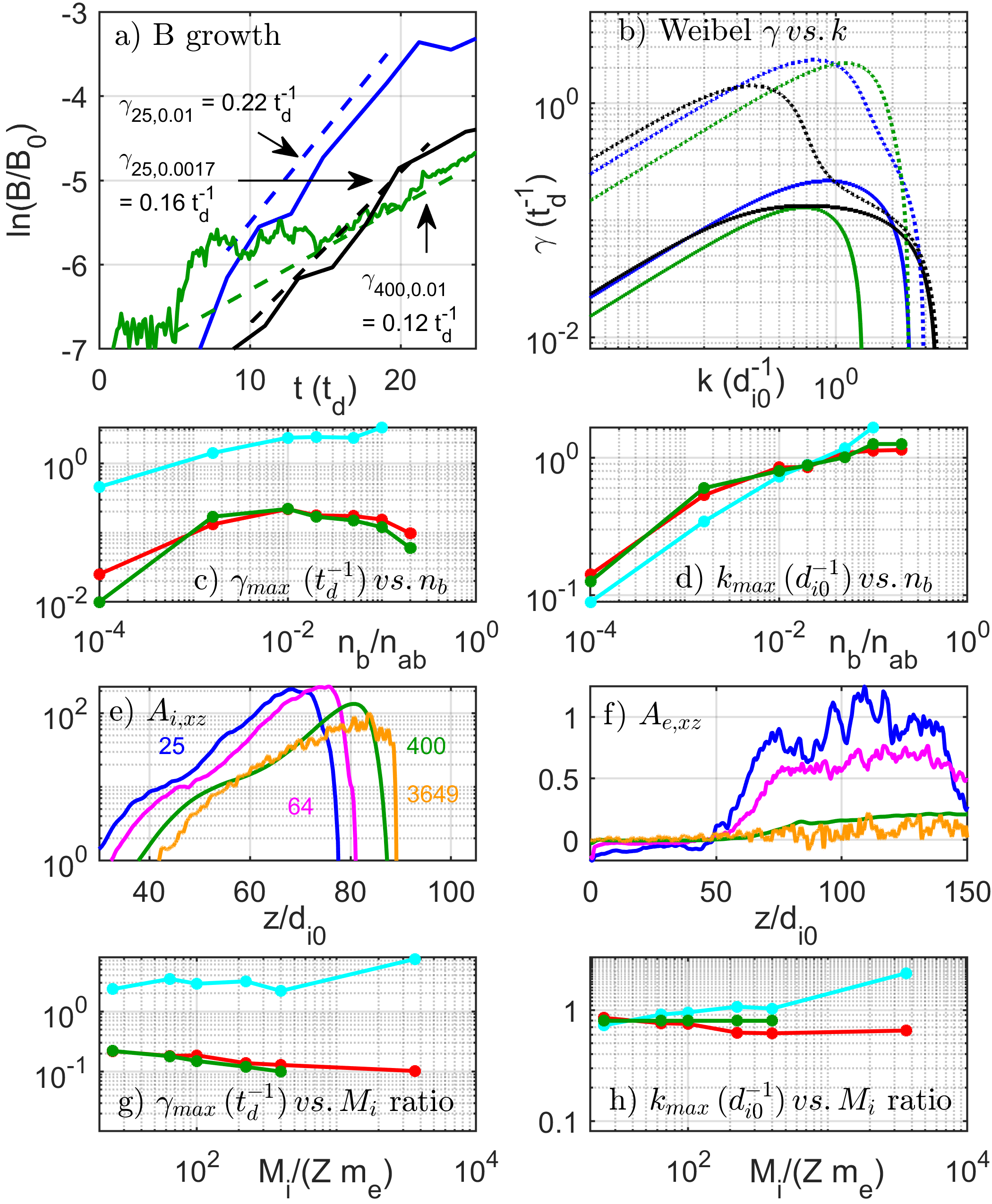}
\caption{(a) Magnetic filamentary growth (peak to peak) vs. time for $n_b/n_{ab} = 0.01$ (blue), $n_b/n_{ab} = 0.0017$ (blue), both at $M_i/(Zm_e) = 25$, and $n_b/n_{ab} = 0.01$, $M_i/(Zm_e) = 400$ (green). Exponential fits in dashed lines: $\gamma_{0.01, 25} = 0.22 t_d^{-1}$, $\gamma_{0.0017, 25} = 0.16 t_d^{-1}$, $\gamma_{0.01, 400} = 0.12 t_d^{-1}$. (b) Weibel dispersion relations obtained by solving Eq. (1) for parameters from each simulation- same colors as (a); solid lines show $\gamma_i$, and dashed lines show $\gamma_{ie}$. (c) $\gamma_{max,i}$ (red) and $\gamma_{ie,max}$ (cyan) vs. $n_b$ at $M_i/(Zm_e) = 25$, with results from simulation (green). (d) Wave number $k$ corresponding to $\gamma_{max,s}$ for each case respectively. (e) $A_i$ measured for 4 mass ratios along $z$ (averaged in $x$) at $t/t_d = 14$, with $M_i/Z m_e= 25$ (blue), $64$ (magenta), green -$400$ (green), and $3600$ (red)- the last simulation was performed in 1-D. (f) $A_e$ measured in the same manner for the same mass ratios. (f,h) Similar quantities to (c,d) but for $M_i/(Zm_e)$ scan at $n_b/n_{ab} = 0.01$.}
\label{fig:3}
\end{figure}

Given $A_i$, we compare B-field growth to the maximum growth rate $\gamma_{max}$ predicted by the theoretical Weibel dispersion relation, which can be written as follows:
\begin{equation}
\label{eq:curlohm}
0 = c^2 k^2 - \omega_k^2 - \sum_{s} \omega_{p,s}^2[A_{s}] - \sum_{s} \omega_{p,s}^2[A_{s} + 1] \xi_s Z(\xi_s),
\end{equation}
where the relationship is summed over all species $s$, $A_{s} = T_{\perp,s}/T_{\parallel,s} -1$, $\xi_s = \omega_k/(kv_{th, \parallel})$, and $Z(\xi_s)$ is the plasma dispersion function \cite{DavidsonPoF1972}. Here $\parallel$ refers to the wavevector direction $\vec{k}$ and $\perp$ the counterstreaming direction ($z$). The ion Weibel regime is characterized by $A_i$ driving the instability, where in electron Weibel, $A_e$ largely determines the instability dispersion relation.


In Fig.\ 3(a), the exponential growth of the magnetic field amplitudes, tracked within the region of maximum $A_i$, is plotted and fitted for $n_b/n_{ab}$= 0.01 (blue) and 0.0017 (black). In Fig.\ 3(b), using the same color coding, we plot the solutions to Eq.\ 1 using parameters ($A_i, A_e, T_{xx,s}$, $n_i$, $n_e$) obtained from simulations; solid lines show solutions to Eq.\ \ref{eq:curlohm} considering only the ion anisotropy ($\gamma_i$), and dashed lines show solutions where both $A_e$ and $A_i$ are used in the dispersion relation solution ($\gamma_{ie}$). Note in Fig.\ 3(f), at $n_b/n_{ab}$= 0.01, $M_i/(Zm_e)=$ 25 (blue), $A_e$ is non-negligible. In Fig.\ 3(c), we compare the maximum dispersion relation growth rates, $\gamma_{i,max}$ (red) and $\gamma_{ie,max}$ (cyan), and the measured simulation growth rates (green) vs. $n_b$. We find good agreement in trend and magnitude between $\gamma_{i,max}$ and the measured growth rates, peaking at $\gamma_{max} = 0.2/t_d$ at $n_b = 0.01$; in contrast, $\gamma_{ie,max}$ overpredicts the growth rate by an order of magnitude, indicating this instability is solely driven by ion anisotropy. In Fig.\ 3(c), $k_{max}$, the $k$ which corresponds to $\gamma_{max}$, also exhibits excellent agreement between the ion-driven solutions and the measured initial wave number. 

Additionally we perform convergence studies of $A_i$, $A_e$, and instability growth with respect to the computational mass ratio, $M_i/(Zm_e)$; here we shrink the box to $L_x = 25 \, d_{i0}$. In Fig.\ 3(e), we find for $n_b/n_{ab}$= 0.01, $A_i$ has a weak mass ratio dependence, dropping from 200 to 130 as $M_i/(Zm_e)$ increases from 25 to 400; this trend is due to perpendicular collisional heating of $T_{xx,i}$ by electrons at the higher mass, where the ion populations remain separated by the same velocity in $v_z$ space, as seen by the green curves in Fig.\ 2(a,b), which show $f(v_z)$ and $f(v_x)$ at $M_i/(Zm_e)= 400$, respectively,  In contrast, the $A_e$ dependence on $M_i/(Zm_e)$ drops dramatically from  $A_e = 1$ to 0.2 over the same interval, and to noise levels at the physical ratio. In Fig.\ 3(a), we include the B-field growth from a simulation at $M_i/(Zm_e)$ = 400 (green), and solutions to Eq.\ 1 in  Fig.\ 3(b) using parameters from this simulation. Similar to our $n_b$ comparison, Fig.\ 3(g,h) compares $\gamma_{max}$ and $k_{max}$ vs. $M_i/(Zm_e)$ from the $\gamma_i(k)$ (red) and $\gamma_{ie}(k)$ (cyan) solutions vs. growth observed in simulations, confirming agreement with an ion-driven Weibel regime, and finding a weak dependence of $\gamma_{max}$ on $M_i/(Zm_e)$, with $\gamma_{phys} = 0.1 t_d^{-1}$.  


Figure 3(c,d,g,h) show excellent agreement with the ion Weibel theory, rather than combined electron-ion Weibel. It is somewhat fortuitous that this results even at low mass ratios. While $A_e$ tends to zero at $M_i/Zm_e$ = 3600, it is not precisely clear why electron-driven Weibel is not observed at the lower mass ratios where $A_e$ persists at appreciable values. Possibly, the instability exists and but saturates far below subsequent ion Weibel growth; alternatively, Ref. \cite{SkoutnevAPJ2019} shows that the electron-type Weibel instability is disrupted by the two-stream instability when $v_{stream} \le 0.2 \,c$; in our simulations $v_{stream} \approx 4\, C_s$, so that at $M_i/(Zm_e)= 25$, $v_{stream} = 0.16\, c$, thus this is also a possible explanation.

\begin{figure}
\includegraphics[width= 8.5cm]{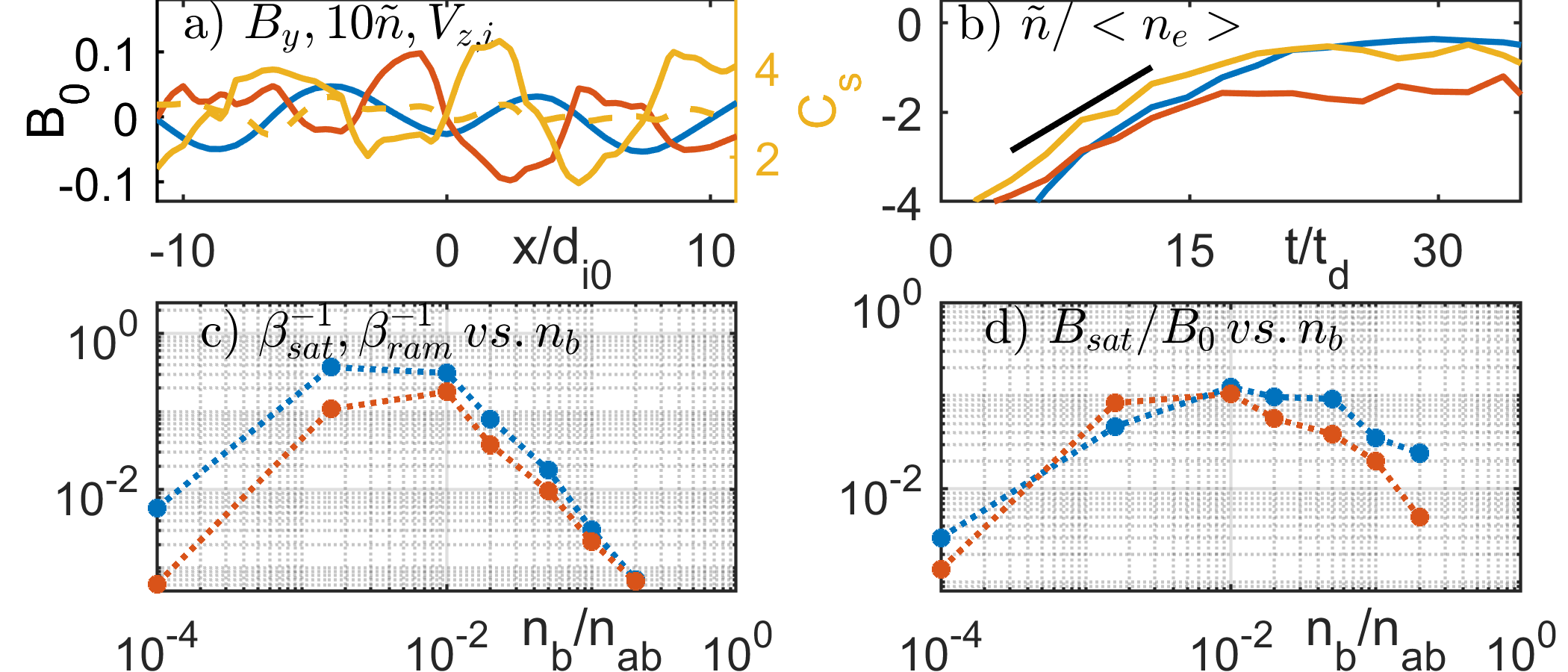}
\caption{(a) 1-D horizontal cut at $z/d_{i0} = 75$ of B-field (blue), 10$\times$ $\tilde{n_e}$ (red), $V_{zi}$ (yellow), and $V_{ze}$ (dashed yellow) at $t/t_d = 20$ for $M_i/(Zm_e) = 25$, $n_b/n_{ab} = 0.01$ (5000 particles at $n_{ab}$). (b) Amplitude of $\tilde{n_e}/\langle n_e\rangle$ (red), $\tilde{V_i}/\langle V_i\rangle$ (yellow), and  $k_x \tilde{B_y}/Ze n_i\langle V_i\rangle$ (blue). (c) Saturation $\beta_{e}^{-1}$ (red) and $\beta_{ram}^{-1}$ (blue) vs. $n_b$. (d) Saturation $B$ (blue) and bounce frequency prediction (red) vs. $n_b$.}
\label{fig:4}
\end{figure}

During filamentation growth, we find that the Weibel instability leads to significant density perturbations in these evolving non-uniform plasmas.
Interestingly, this contrasts with the standard Weibel linear theory for two symmetric counterstreaming beams, in which no density perturbations are predicted \cite{DavidsonPoF1972}; however, our scenario differs from Davidson et al. due to the large density gradient in the streaming direction. In Fig.\ 4(a), horizontal cuts through the peak anisotropy of the magnetic field, $10\times$ $\tilde{n_e}$, $V_{zi}$, and $V_{ze}$ are plotted at $t/t_d = 20$ for the $M_i/(Zm_e) = 25$, $n_b/n_{ab} = 0.01$ simulation; here $\tilde{n_e} = n_e - \langle n_e \rangle$, where $\langle n_e \rangle$ defines the $x$ averaged density at a particular $z$. Seen in Fig.\ 4(b), inspecting the growth of $\tilde{n_e}/\langle n_e\rangle$ (red), $\tilde{V_{zi}}/\langle V_{zi}\rangle$ (yellow) and $k_x \tilde{B_y}/Ze n_i\langle V_{zi}\rangle$ (blue), we find the density perturbations grow linearly with the instability, where $\tilde{n_e}$ saturates at $\approx$ 0.005 $n_{ab}$. $\tilde{B_y}$ has been normalized via Ampere's law, assuming $j \approx Ze n_i\tilde{V_{zi}}$, and $i k_x B_y \approx  \mu_0  j_z$.


Investigating the instability saturation, Fig.\ 4(c) shows the saturated value of $\beta_{e}^{-1} = B^2/2\mu_0n_eT_e$ and $\beta_{ram}^{-1} = B^2/\mu_0n_i M_i V_{z,i}^2$, the magnetic pressure over the local plasma pressure and ion flow energy, respectively, versus $n_b$. Notably, the magnetic pressure can reach over 10\% of the plasma pressure. Inspecting the saturation of the B-field, we find agreement with the argument in Ref.\ \cite{DavidsonPoF1972}, where saturation occurs when the ion bounce frequency $\omega_{b,i} = \sqrt{kV_yB}$ matches the instability growth rate $\gamma_w$. From this argument, we find $B = \gamma_w^2 k V_y$. Given the observed $\gamma_w$ and $k$ values, we calculate the expected saturation values of $B$ (red), which in Fig.\ 4(d) are plotted versus the maximum B-field (blue) observed via simulation, demonstrating quantitative agreement over $n_b$. We predict B-field saturation over 0.05 $B_0$, or 50 T, when $n_b/n_{ab} = 0.0017 -0.05$.
%



In conclusion, we observe in a NIF-like parameter regime that the ion-type Weibel instability generates $\sim$100 T magnetic filaments at a wavelength of 100 - 250 $\mu m$ with associated density oscillations within the corona of a single ablated plasma, provided there exists a background plasma population. 
This instability could explain several experiments where filamentation within a single plume expansion is observed and not yet understood \cite{ManuelPoP2013, RyggScience2008, RosenbergPRL2015, GaoPRL2015}. Most significantly, this process is important to indirect drive inertial confinement experiments where hohlraums can contain a helium fill gas at $5\times10^{24}$ - $2\times10^{26}$ m$^{-3}$ (0.001-0.05 $n_{ab}$), the optimal density range for this instability to grow. 

%
\begin{acknowledgments}
Simulations were conducted on the Titan supercomputer at the Oak Ridge Leadership Computing Facility at the Oak Ridge National Laboratory, supported by the Office of Science of the DOE under Contract No. DE-AC05-00OR22725. J. M. was supported by the DOD through the NDSEG Program, 32 CFR 168a. This research was also supported by the DOE under Contracts No. DE-SC0006670 and No. DE-SC0016249. We thank D. Strozzi for the helpful discussions.
\end{acknowledgments}



\bibliographystyle{apsrev4-1}

\bibliography{../citeulike_short}

\end{document}